\documentclass[epjCONF]{svjour} 
\usepackage[varg]{txfonts} 
\usepackage[latin1]{inputenc}
\usepackage{graphicx}

\def\df{{\sc df}}
\def\bolJ{\mathbf{J}}

\def\bolw{\mathbf{w}}
\def\vsol{\mathbf{v_\odot}}
\def\kms{\,\mathrm{km\,s}^{-1}}
\def\bolth{\mbox{\boldmath$\theta$}}
\def\bolom{\mbox{\boldmath$\Omega$}}

\session-title{Assembling the Puzzle of the Milky Way}
\begin{document}
\title{Dynamical models of the Galaxy} \author{Paul J. McMillan
  \thanks{\email{p.mcmillan1@physics.ox.ac.uk}}} \institute{Rudolf
  Peierls Centre for Theoretical Physics, 1 Keble Road, Oxford, OX1
  3NP, UK}

\abstract{ I discuss the importance of dynamical models for exploiting
  survey data, focusing on the advantages of ``torus'' models. I
  summarize a number of applications of these models to the study of
  the Milky Way, including the determination of the peculiar Solar
  velocity and investigation of the Hyades moving group.}

\maketitle

\section{Introduction}
\label{intro}
Current studies of the structure of the Milky Way are dominated by a
series of major observational programs running from ESA's Hipparcos
mission \cite{Hipparcos}, through major photometric surveys such as
the SDSS \cite{Sloan}, spectroscopic surveys such as RAVE \cite{RAVE},
to ESA's scheduled Gaia mission \cite{Gaia}, which aims to return
photometric and astrometric data for $10^9$ stars and low-dispersion
spectra for $\sim10^8$ stars.

Turning these data-sets into a consistent picture of the current
structure and the assembly of the whole Galaxy, including the
dark-matter content, is an ambitious and important goal. It is likely
to be impossible without sufficiently sophisticated models that can be
used to interpret the data (for example compensating for the
observational biases of the various surveys).

Models of the gross structure of the Galaxy have been produced with
varying levels of complexity. Mass models
\cite{WDJJB98:mass,PJM11:mass} simply give the density distribution
of the various components of the Galaxy, and thus the Galactic
potential. Kinematic models, such as those produced by
\textsc{galaxia} \cite{Galaxia}, specify the density and velocity
distributions of the luminous components of the Galaxy, but do not
consider the question of whether these are consistent with a steady
state in any Galactic potential. The Besan\c{c}on Galaxy model
\cite{Roea03} is primarily a kinematic model (in that it is not
constrained by Newton's laws of motion on large scales), with a
dynamical element used to determine the vertical structure of the
disc. Fully dynamical models \cite{PJMJJB11:LOS} have a distribution
of stars in phase-space which is made from phase-mixed orbits in a given 
Galactic potential, and 
therefore a distribution function (\df) which is a function of the integrals 
of motion, which by Jeans theorem \cite{Je15} means it is in a steady-state.


\section{Benefits of dynamical models}
\label{sec:benefits}

A previous paper \cite{PJMJJB11:LOS} provides a detailed discussion
of the benefits of dynamical models. Here I just sum up the main
points.

The most obvious advantage of dynamical models is that, unlike
kinematic models, they allow us to deduce the gravitational potential
of the Galaxy. Existing mass models were fit to observations under the
assumption of near circular orbits for various tracers, which is only
suitable for a small subset of astrophysical objects, found close to
the Galactic plane. Exploiting richer data-sets requires far more
careful modelling. This will allow us to infer the distribution of
dark matter in the Galaxy, under the assumption that the Galaxy is
approximately in a steady state.

A second, complementary, advantage is that dynamical models allow us
to connect objects that we can observe to objects those we can not.
This means we can use observations in the solar
neighbourhood to learn about the structure of the Galaxy at large.
For example \cite{MayB} showed that if the stellar halo were in virial
equilibrium, more than half the stars of the stellar halo would be on
orbits that bring them through the solar neighbourhood. 

An additional advantage of dynamical models is that the associated
\df\ depends only on the three integrals of
motion (either explicitly or implicitly), as opposed to the full six
dimensions of phase-space.

\section{Torus models}
Dynamical modelling has been dominated by Schwarzschild modelling
\cite{SchwarzI}, especially for analysing the dynamics of early-type
galaxies. This technique involves first integrating a number of orbits
in the adopted gravitational potential and then seeking weights for
these orbits such that the weighted sum of the orbits reproduces the
observational data. More recently, the ``made-to-measure'' (M2M)
technique introduced by \cite{SyerT} has been used to produce $N$-body
models which can be fitted in a broadly similar way
\cite{deLorenzi,Bissantz04}, though in this case the particle
weights (which are effectively the orbit weights) are determined
``on-the-fly'', rather than after the orbit has been integrated.

\begin{figure*}
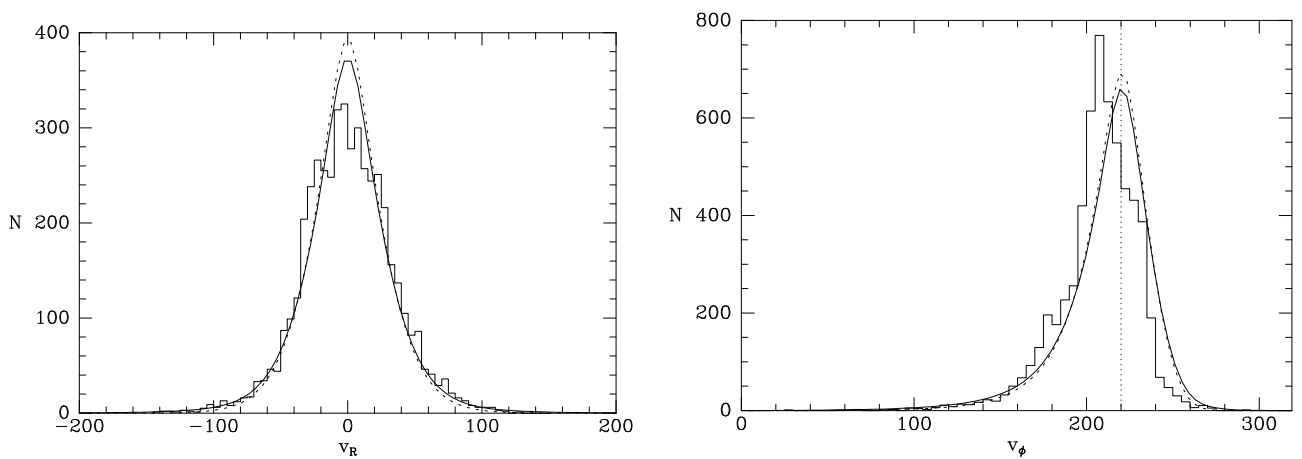

  \centerline{\hfil
    \resizebox{82mm}{!}{\includegraphics{McMillan_P_fig1a.eps}}\hspace{4mm}
    \resizebox{82mm}{!}{\includegraphics{McMillan_P_fig1b.eps}}}
  \caption{Distribution of stars in $v_R$ (left) and $v_\phi$ (right)
    in the Solar neighbourhood, data from the Geneva-Copenhagen survey
    assuming the value of $\vsol$ given in eq.~\ref{eq:vsol} 
    (histogram), and model from \cite{JJB10} which
    is a single \df\ $f(\bolJ)$ which must fit \emph{both} the $v_R$
    and $v_\phi$ distributions (curves, solid curve is from the full
    model, dashed curve is thin disk contribution only). The dotted
    vertical line in the $v_\phi$ plot is the assumed circular speed
    at the Sun. Clearly the $v_\phi$ distributions are significantly
    offset from one another. The only way to bring the data and model
    to reasonable agreement is to apply a correction of $\sim7\kms$
    to $V_\odot$. Figure reproduced from \cite{JJB10} with
    permission.}
  \label{fig:vsol} 
\end{figure*}

Both Schwarzschild and M2M models have \df s which are implicitly
function of the integrals of motion, as they are constructed from
phase-averaged orbits. An alternative modelling strategy is to produce
models which are \emph{explicitly} functions of the integrals of
motion. It is not possible to produce plausible models of the Galactic
disc with \df s which are functions of the classical integrals,
$f(E,L_z)$, as these models have equal velocity dispersions in the radial and
vertical directions, which is not the case in the Solar neighbourhood
\cite{AumerB09}. However, one can instead use the three orbital
``actions'', $J_i$, as the integrals of motion and use analytic \df s,
$f(\bolJ)$, which are appropriate for the Galactic disc
(e.g. \cite{JJB10}).

The three actions $J_i$ and three conjugate angles coordinates
$\theta_i$ provide canonical coordinates for six-dimensional phase
space \cite{BT08}.  The conventional phase space coordinates
$\mathbf{w}\equiv({\bf x},{\bf v})$ are $2\pi$-periodic in the
angles. The actions are conserved quantities for any orbit, and the
angles increase linearly with time:
\begin{equation}
  \bolth(t) = \bolth(0) + \bolom(\bolJ) t,
\end{equation}
where the components of $\bolom$ are the orbital frequencies.

The major obstacle to using \df s of the form $f(\bolJ)$ is that the
relationship between phase space coordinates $\mathbf{w}$ and
$\bolJ,\bolth$ is only known analytically for a very limited set of
gravitational potentials, none of which provides a reasonable
approximation to the Galactic potential. There are two available
approaches for determining the relationship between $\mathbf{w}$ and
$\bolJ$:
\begin{itemize}
\item The adiabatic approximation, in which motion in the radial and
  vertical directions are largely decoupled \cite{JJB10,JJBPJM11}.
\item Torus modelling, in which the relationship between
  $\bolJ,\bolth$ and $\bolw$ in an isochrone potential (for which it
  is known analytically \cite{BT08}) is numerically distorted to fit
  the potential of interest \cite{McGJJB90,KaJJB94,PJMJJB08}.
\end{itemize}
Comparison of these two approaches has shown that for most purposes
they agree to reasonable accuracy up to as far as
$\sim2.5\,\mathrm{kpc}$ from the Galactic plane \cite{JJBPJM11}. The
adiabatic approximation has the advantage that it does not require
specialised torus-fitting computer code, and can straightforwardly
determine the value of $\bolJ$ for a given $\bolw$.  Torus modelling
finds all of the values of $\bolw$ associated with a given $\bolJ$,
but can only find $\bolJ$ given $\bolw$ as an iterative process. Torus
modelling has the advantages that it can tell us about the coupling
between different components of motion (e.g. the tilt of the velocity
ellipsoid \cite{JJBPJM11}), and that it allows us to find the angle
variables $\bolth$.

\section{The local standard of rest}

As mentioned in Section~\ref{sec:benefits}, one major advantage of
using dynamical models is that the dimensionality of the models is
reduced by the assumption that the Galaxy is made up of phase mixed orbits --
$f(\bolJ)$ depends on only three actions as opposed to the six
dimensions of $f(\bolw)$. The value of this simplification of the
model is well illustrated by a recent revision in the peculiar Solar
velocity relative to the local standard of rest, $\vsol$.

The value of $\vsol$ was found by \cite{WDJJB98:LSR} using
observations by Hipparcos. The components of velocity towards the
Galactic centre ($U$), in the direction of Galactic rotation ($V$) and
towards the north Galactic pole ($W$) were analysed separately. The
$V$-component of the Solar velocity is the most difficult to determine
as asymmetric drift \cite{BT08} means that average stellar velocity lags the
circular velocity. Stromberg's equation was used by \cite{WDJJB98:LSR}
to extrapolate from the observed populations (separated by colour) to
a hypothetical population with zero velocity dispersion, which would
have zero asymmetric drift. The value of $\vsol$ found,
\begin{equation} \label{eq:vsol}
U_\odot,V_\odot,W_\odot = (10.00\pm0.36,\,5.25\pm0.62,\,7.17\pm0.38)\kms,
\end{equation}
was the widely accepted value for over a decade.

Figure~\ref{fig:vsol} shows the distribution of stars in $v_\phi$ and
$v_R$ in the Solar neighbourhood, determined from Hipparcos
observations and the Geneva-Copenhagen survey \cite{GCS04} assuming
this value of $\vsol$ (histogram), and the best
fitting analytic \df \ $f(\bolJ)$ from \cite{JJB10} (solid curve). 
This gives new insight because the 
distributions in $U$ and $V$ are not considered as if they were
independent, but instead it is recognised that a single dynamically
consistent \df\ must fit both. The only physically plausible way to 
bring the model and data $v_\phi$ distributions into agreement is to 
apply a correction of $\sim7\kms$ to the value of $V_\odot$ (altering 
the circular velocity moves both distributions and has a negligible 
effect). This is $\sim11\sigma$ from the widely accepted value!
This contention, that the previously
assumed value of $V_\odot$ must be in error, has since been supported
by analysis of Galactic maser sources with measured parallaxes
\cite{PJMJJB10}, and explained as an error in the application of
Stromberg's equation, associated with the metallicity gradient in the
disc \cite{SBD10}.

\section{Beyond steady-state models}
Axisymmetric models with \df s of the form $f(\bolJ)$ cannot fully
describe the Galaxy, as it is not in fact in a steady state, but they are an
important first step towards interpreting observations of our Galaxy,
and it is likely to be very fruitful to study observational data for
structures that cannot be explained by these models. We can then look
for explanations of these structures as signatures of other features,
such as the Galactic bar, spiral or warp, or matter that has been accreted. 
Some examples
of using torus models to explore these signatures already exist.

\subsection{Signatures of accreted satellites}

\begin{figure}
  \centerline{\hfil
    \resizebox{41mm}{!}{\includegraphics{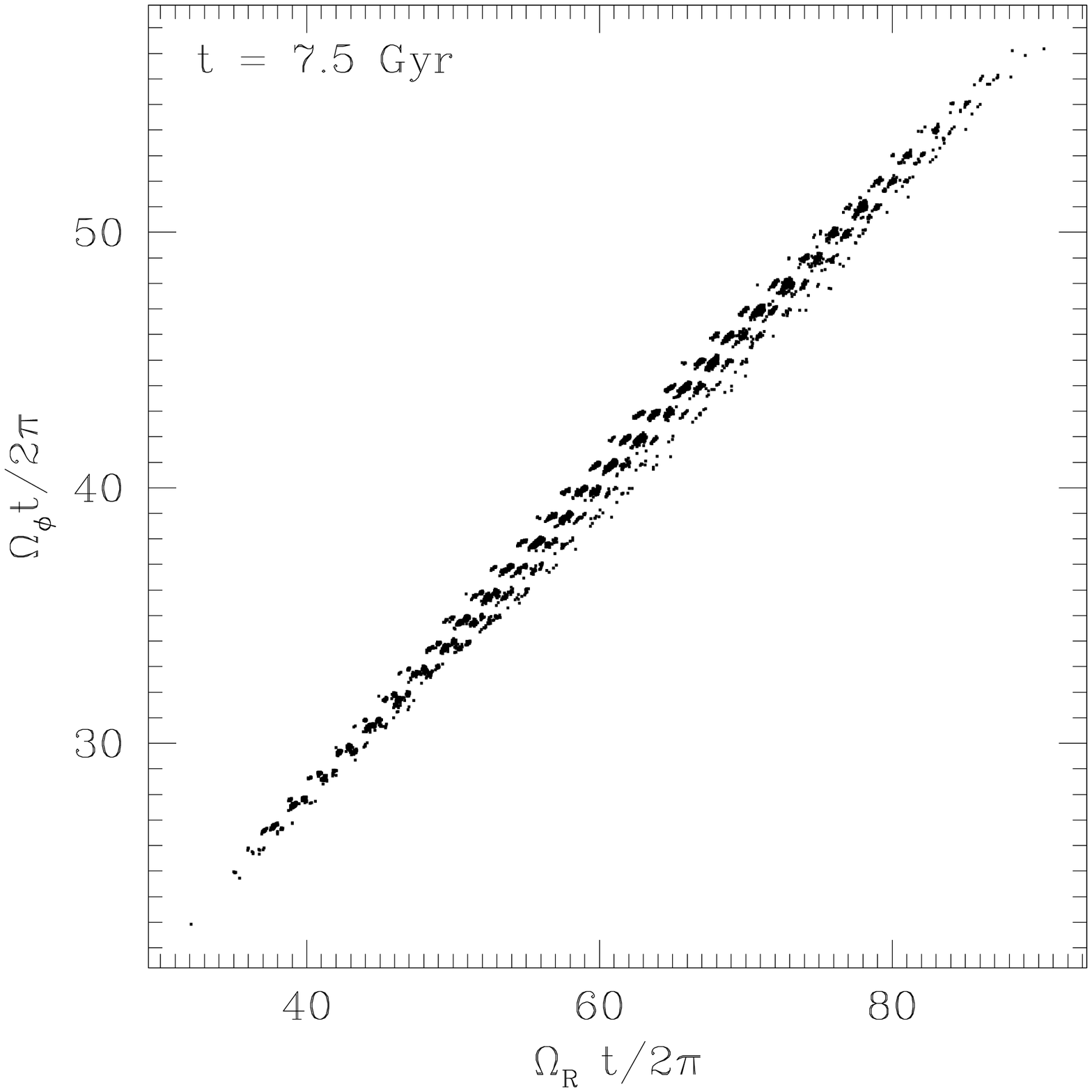}}\hspace{4mm}
    \resizebox{41mm}{!}{\includegraphics{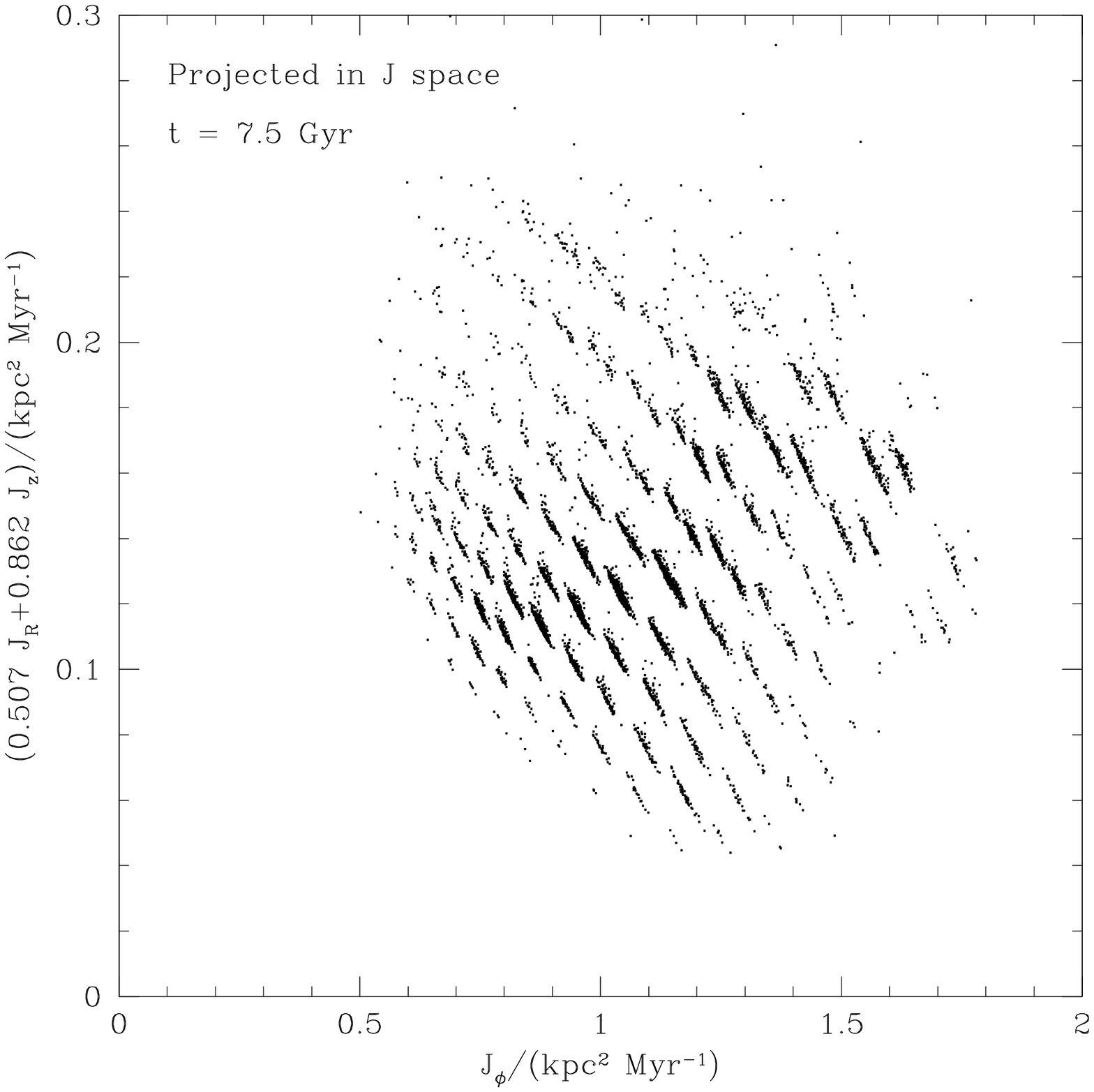}}}
  \caption{Frequencies $\Omega_r$ \& $\Omega_\phi$ (left) and a
    convenient projection in action space (right) for stars in the
    solar neighbourhood from a simulation of the accretion of a
    satellite galaxy, taken $t=7.5\,\mathrm{Gyr}$ after the satellite
    was accreted. The stars form patches in frequency space with
    spacing $\Delta\Omega_\phi$ between patches such that
    $\Delta\Omega_\phi t/2\pi\sim1$ (the spacing in $\Omega_r$ is not
    as simple because a two ranges of $\theta_r$ correspond to the
    solar neighbourhood).  The same separation into discrete clumps is
    visible in action because $\bolom=\bolom(\bolJ)$. These figures
    are taken from \cite{PJMJJB08}.}
  \label{fig:patches}
\end{figure}
The appearance in angle-action coordinates of an accreted satellite
galaxy was explored by \cite{PJMJJB08}. Long after phase mixing has
rendered an accreted satellite indistinguishable from the background
population in position, there is a strong relationship between the
stars' positions and their orbital frequencies (because all the stars were
once collected in the same small volume, when they were part of the
satellite). This means that a sample of these stars taken from a
finite volume is only found in certain small volumes in frequency
space.

In Fig~\ref{fig:patches} I show figures from \cite{PJMJJB08} of the
frequencies and actions of stars in a finite volume about the solar
position in a simulation of the disruption of a satellite galaxy. In
both cases the figure shows the distribution $7.5\,\mathrm{Gyr}$ after
the satellite was disrupted. The distribution in frequency is clearly
divided into finite ``patches'', and the distribution in action is
even more cleanly divided, because $\bolom=\bolom(\bolJ)$ and it
provides a more convenient projection of the distribution.

By considering the spacing between the ``patches'' in frequency space,
along with the angles of the individual stars, \cite{PJMJJB08} showed
that it was even possible to determine with high accuracy the time
since the satellite was disrupted. Similar techniques could even be
used to determine the potential of the Galaxy (as the potential must
allow the stars to all have come from the same initial
satellite). Similar work has been carried out by other authors showing
that using the orbital frequencies alone one can achieve some of these
objectives, even for cosmological simulations with numerous accreting
satellites and a non-static background potential \cite{GoHe10}.

\subsection{Signatures of Lindblad resonances}

\begin{figure}
  \centerline{\hfil
    \resizebox{\hsize}{!}{\includegraphics[angle=270]{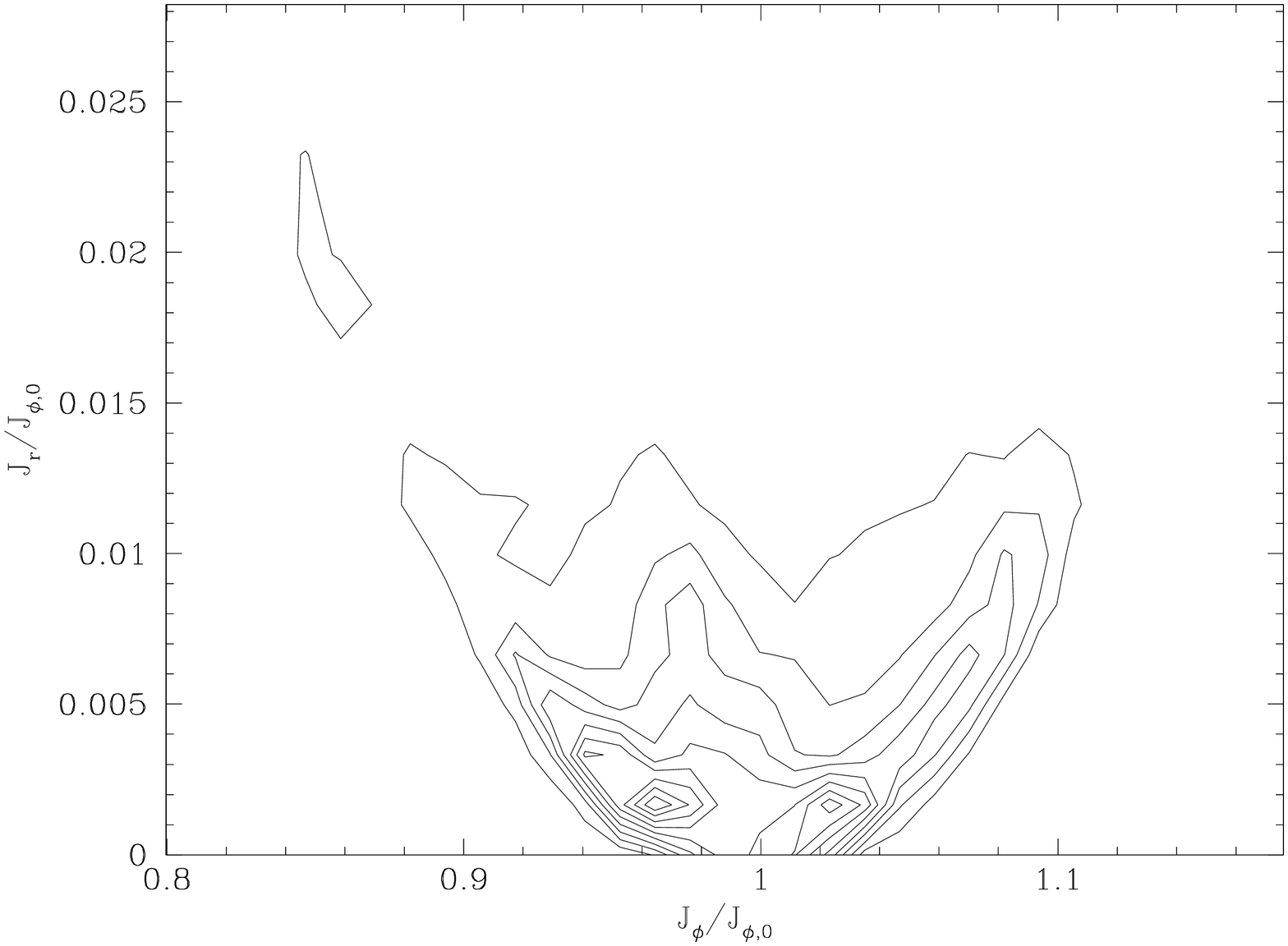}}}\vspace{-4mm}
  \centerline{\hfil
    \resizebox{\hsize}{!}{\includegraphics[angle=270]{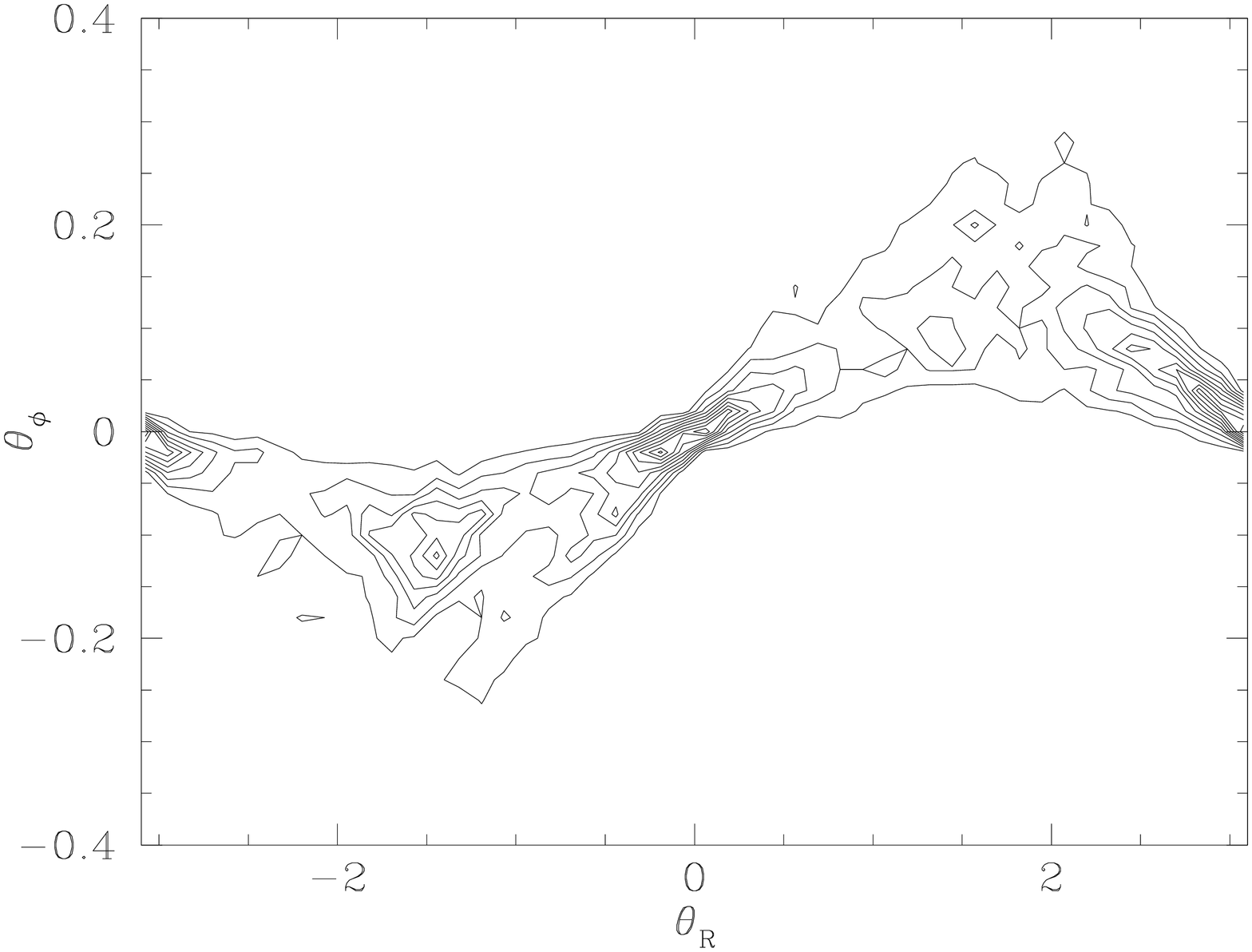}}}
  \caption{Distribution of stars in action (upper) and angle (lower)
    in the Solar neighbourhood with positions and velocities given by
    the Geneva Copenhagen survey. The gross structures of the
    distributions are due to selection effects. The Hyades moving
    group can be seen as an overdensity in action that is spread out
    in $J_r$ at around $J_\phi=0.97$, tending towards slightly lower
    $J_\phi$ with increasing $J_r$ (consistent with a resonance line
    in $\bolom$), and as an overdensity in angle around
    $\theta_r=-\pi/2$. The expected angle relation for stars at
    resonance would produce overdensities that tend to increased
    $\theta_r$ with increased $\theta_\phi$ for inner Lindblad
    resonances, and to decreased $\theta_r$ with increased
    $\theta_\phi$ for outer Lindblad resonances. However, selection
    effects reshape overdensities in angle space quite significantly.
    Figures are adapted from \cite{PJM11:Res}
  }
  \label{fig:hyades} 
\end{figure}

A recent study \cite{Se10} showed that in addition to having a
relationship of the form $l\Omega_r+m\Omega_\phi \simeq const$, stars
that have recently been trapped at a resonance with a perturber also
follow a relationship of the form $l\theta_r+m\theta_\phi \simeq
const$, where in both cases $l$ and $m$ are integers, with the
perturbation having $m$-fold symmetry and $l$ being $-1$ for an inner
Lindblad resonance and $+1$ for an outer Lindblad resonance. The
Hyades moving group, which is a very strong feature of the local
velocity distribution \cite{WD98}, lies around a straight line in the
$J_\phi,J_r$ plane (Fig~\ref{fig:hyades}), of the sort associated with
the condition $l\Omega_r+m\Omega_\phi \simeq const$ (a resonance line
in action space).  This finding is indeed consistent with a resonance
line for the both the cases $l=\pm1$ and a range of values of $m$
(with the exact details of the resonance lines depending quite
sensitively on the details of the Galactic potential assumed).

It was claimed by \cite{Se10} that the distribution of stars in angle
coordinates clearly indicated that the Hyades were associated with an
inner Lindblad resonance. This is because, in angle space, the stars
of the Hyades moving group are associated with an overdensity in the
quantity $-\theta_r+m\theta_\phi$ for various values of
$m$, and this overdensity did not appear to significantly shift in
position as a function of $J_r$.

In \cite{PJM11:Res} I compared the observed structure to torus models,
and demonstrated the significant and non-intuitive impact of selection
effects. In Fig~\ref{fig:hyades} I show the distribution of Solar
neighbourhood stars in the $\theta_r,\theta_\phi$ plane. Selection
effects are responsible for the overall structure of the density
distribution, most notably the high densities around $\theta_\phi=0$,
$\theta_r=0$ or $\pm\pi$ and the near absence of stars with
$\theta_r<0$, $\theta_\phi>0$ or $\theta_r>0$, $\theta_\phi<0$. A less
obvious selection effect brought to light by torus models is that
stars with a given value of $\bolJ$, found in the Solar neighbourhood
are also very strongly restricted in their possible range of $\bolth$.

Using these models I was able to show that observed overdensities in
$\theta_r+m\theta_\phi$ (which should correspond to
outer Lindblad resonances) were not the result of selection effects
(as claimed by \cite{Se10}).  Also, these selection effects mean that
the stars associated with the resonance line in action space have to lie
near certain lines in angle space, otherwise they will not be observed in
the Solar neighbourhood. It is the interplay of the resonant
conditions on $\bolJ$ and $\bolth$, and the fact that the stars lie in
the Solar neighbourhood that determines the distribution in both angle
\emph{and} action, and considering either distribution independently
of the other can lead to false conclusions. The examination of torus
models which included resonant components (with resonant conditions on
both $\bolJ$ and $\bolth$) indicated that the Hyades overdensity was
consistent with stars trapped at \emph{either} an inner \emph{or} outer
Lindblad resonance.

\section*{Acknowledgments}
I am grateful to James Binney who was co-author or author on many of
papers used as examples in this article, and provided helpful
suggestions on presentation.  This work is supported by a grant
from the Science and Technology Facilities Council.

\end{document}